\documentclass[english]{revtex4}
\usepackage[T1]{fontenc}
\usepackage[latin1]{inputenc}
\usepackage{graphicx}

\makeatletter

\usepackage{babel}
\makeatother
\begin{document}

\title{Spinning binary waveforms via PN expansion: Equal-mass case}

\author{Dong-Hoon Kim}

\affiliation{Center for Quantum Spacetime, \#310 Ricci Annex Hall, Sogang University,
Shinsu-dong Mapogu Seoul 121-742, Korea \\
}

\begin{abstract}
Complete expressions of time-domain gravitational waveforms for spinning
binary inspirals via the post-Newtonian (PN) approximation would require
determination of the phase, amplitude, inclination angle, precession
phase and spin vectors as well as the knowledge of the order coefficients
for the PN expansion terms. These quantities are determined by solving
simultaneously the spin-precession equations, the evolution equation
for the Newtonian angular momentum, and the equation for the orbital
frequency. For the spinning binaries with equal masses, determination
of these quantities can be done fully analytically, by taking advantage
of the equal mass symmetry, therefore by simplifying those equations
to solve. We provide the analytical results through 1.5 PN order which
includes spin-orbit interactions. 
\end{abstract}
\maketitle
\emph{Arun et. al}\citet{Arun et al:2009} provides the {}``Ready-to-use''
time-domain gravitational waveforms for spinning binary inspirals
in Post-Newtonian expansion to 1.5 order. Their formulations, however,
have yet to be further specified, depending on the configurations
of binaries, such as the mass ratio and spin alignment with the Newtonian
angular momentum. In order to make use of their theoretical waveforms
in designing our wave templates via the IIR method, it is necessary
to solve the equations governing these configurations and to fully
specify the waveforms with all known parameters. 

The set of equations to solve are the following: \\
(i) the spin-precession equations

\begin{eqnarray}
\mathbf{\dot{S}}_{1} & = & \mathbf{\Omega}_{1}\times\mathbf{S}_{1},\label{eq:1}\\
\mathbf{\dot{S}}_{2} & = & \mathbf{\Omega}_{2}\times\mathbf{S}_{2},\label{eq:2}\end{eqnarray}
 where at 1.5 PN order\begin{equation}
\mathbf{\Omega}_{1,2}=M^{2/3}\omega_{\mathrm{orb}}^{5/3}\left(\frac{3}{4}+\frac{\nu}{2}\mp\frac{3}{4}\delta\right)\hat{\mathbf{L}}_{N},\label{eq:3}\end{equation}
with \begin{eqnarray}
M & = & M_{1}+M_{2},\label{eq:4}\\
\nu & = & \frac{M_{1}M_{2}}{M^{2}},\label{eq:5}\\
\delta & = & \frac{M_{1}-M_{2}}{M}.\label{eq:6}\end{eqnarray}
(ii) the evolution equation for the Newtonian angular momentum\begin{equation}
\dot{\hat{\mathbf{L}}}_{N}=-\frac{v}{\nu M^{2}}\left(\mathbf{\dot{S}}_{1}+\mathbf{\dot{S}}_{2}\right),\label{eq:7}\end{equation}
where\begin{equation}
v\equiv\left(M\omega_{\mathrm{orb}}\right)^{1/3}.\label{eq:8}\end{equation}
(iii) the equation for the orbital frequency\begin{eqnarray}
\frac{\dot{\omega}_{\mathrm{orb}}}{\omega_{\mathrm{orb}}^{2}} & = & \frac{96}{5}\nu v^{5}\left\{ 1-\left(\frac{743}{336}+\frac{11}{4}\nu\right)v^{2}\right.\nonumber \\
 &  & \left.+\left[\left(\frac{19}{3}\nu-\frac{113}{12}\right)\mathbf{\chi}_{s}\cdot\hat{\mathbf{L}}_{N}-\frac{113}{12}\delta\mathbf{\mathbf{\chi}}_{a}\cdot\hat{\mathbf{L}}_{N}\right]v^{3}+4\pi v^{3}\right\} ,\label{eq:9}\end{eqnarray}
 where\begin{eqnarray}
\mathbf{\mathbf{\chi}}_{s} & = & \frac{1}{2}\left(\mathbf{\mathbf{\chi}}_{1}+\mathbf{\mathbf{\chi}}_{2}\right),\label{eq:10}\\
\mathbf{\chi}_{a} & = & \frac{1}{2}\left(\mathbf{\chi}_{1}-\mathbf{\chi}_{2}\right)\label{eq:11}\end{eqnarray}
with the normalized spin vectors\begin{equation}
\mathbf{\chi}_{n}=\frac{\mathbf{S}_{n}}{M_{n}^{2}},\;\;\;\;\; n=1,\,2,\label{eq:12}\end{equation}
so that $\left|\mathbf{\chi}_{n}\right|\le1$ for objects that obey
the Kerr bound on rotational angular momentum.

\section{Specifying the waveforms for equal-mass spinning binaries}

\subsection{Configurations for the equal-mass case}

One interesting case might be a spinning binary with equal masses,
where one can take advantage of the relatively simple spinning configurations
due to mass symmetry, thus can analyze the spin effects on the binary
such as precession and change in wave frequency more easily. For an
equal-mass binary, one has\begin{equation}
M_{1}=M_{2}=\frac{M}{2},\label{eq:13}\end{equation}
therefore\begin{eqnarray}
\nu & = & \frac{1}{4},\label{eq:14}\\
\delta & = & 0,\label{eq:15}\end{eqnarray}
and\begin{equation}
\mathbf{\chi}_{n}=\frac{4\mathbf{S}_{n}}{M^{2}},\;\;\;\;\; n=1,\,2,\label{eq:16}\end{equation}
thus,\begin{eqnarray}
\mathbf{\mathbf{\chi}}_{s} & = & \frac{2\mathbf{S}}{M^{2}},\label{eq:17}\\
\mathbf{\mathbf{\chi}}_{a} & = & \frac{2\bar{\mathbf{S}}}{M^{2}},\label{eq:17-1}\end{eqnarray}
 with the definitions\begin{eqnarray}
\mathbf{S} & \equiv & \mathbf{S}_{1}+\mathbf{S}_{2},\label{eq:18}\\
\bar{\mathbf{S}} & \equiv & \mathbf{S}_{1}-\mathbf{S}_{2}.\label{eq:18-1}\end{eqnarray}

With this simplification, one reduces the set of equations (i), (ii)
and (iii) above to the following:\\
(i') the spin-precession equations\begin{eqnarray}
\mathbf{\dot{S}}_{1} & = & \mathbf{\Omega}\times\mathbf{S}_{1},\label{eq:19}\\
\mathbf{\dot{S}}_{2} & = & \mathbf{\Omega}\times\mathbf{S}_{2},\label{eq:20}\end{eqnarray}
with\begin{equation}
\mathbf{\Omega}=\frac{7}{8}M^{2/3}\omega_{\mathrm{orb}}^{5/3}\hat{\mathbf{L}}_{N}.\label{eq:21}\end{equation}
Further, we may combine Eqs.~(\ref{eq:19}) and (\ref{eq:20}) via
Eq.~(\ref{eq:18}), \begin{equation}
\mathbf{\dot{S}}=\mathbf{\Omega}\times\mathbf{S},\label{eq:22}\end{equation}
which will be used throughout the rest of the analysis instead of
Eqs.~(\ref{eq:19}) and (\ref{eq:20}).\\
(ii') the evolution equation for the Newtonian angular momentum\begin{equation}
\dot{\hat{\mathbf{L}}}_{N}=-\frac{4v}{M^{2}}\dot{\mathbf{S}}\label{eq:23}\end{equation}
via Eq.~(\ref{eq:18}).\\
(iii') the equation for the orbital frequency\begin{equation}
\frac{\dot{\omega}_{\mathrm{orb}}}{\omega_{\mathrm{orb}}^{2}}=\frac{24}{5}v^{5}\left[1-\frac{487}{168}v^{2}+\left(4\pi-\frac{47}{3}\frac{\mathbf{S}\cdot\hat{\mathbf{L}}_{N}}{M^{2}}\right)v^{3}\right].\label{eq:24}\end{equation}

\subsection{Solving the configuration-equations in the limit $S\ll L$}

In order to solve the equations above effectively, one needs to prescribe
the time-varying coordinates with respect to the fixed Cartesian coordinates,
in which the Newtonian angular momentum becomes the same as that of
a non-spinning binary. One such kind of the frame of coordinates may
be written as\begin{equation}
\left[\begin{array}{c}
\vec{e}_{x}(t)\\
\vec{e}_{y}(t)\\
\vec{e}_{z}(t)\end{array}\right]=\left[\begin{array}{ccc}
\cos\iota\cos\alpha & \cos\iota\sin\alpha & -\sin\iota\\
-\sin\alpha & \cos\alpha & 0\\
\sin\iota\cos\alpha & \sin\iota\sin\alpha & \cos\iota\end{array}\right]\left[\begin{array}{c}
\vec{e}_{x0}\\
\vec{e}_{y0}\\
\vec{e}_{z0}\end{array}\right],\label{eq:25}\end{equation}
where $\iota$ and $\alpha$ represent the angle of inclination (due
to precession) and the phase of precession, respectively %
\footnote{Ref.\citet{Arun et al:2009} has a slightly different prescription
for the time-varying coordinates.%
}. 

\includegraphics[bb=0bp 275bp 595bp 842bp,clip,scale=0.6]{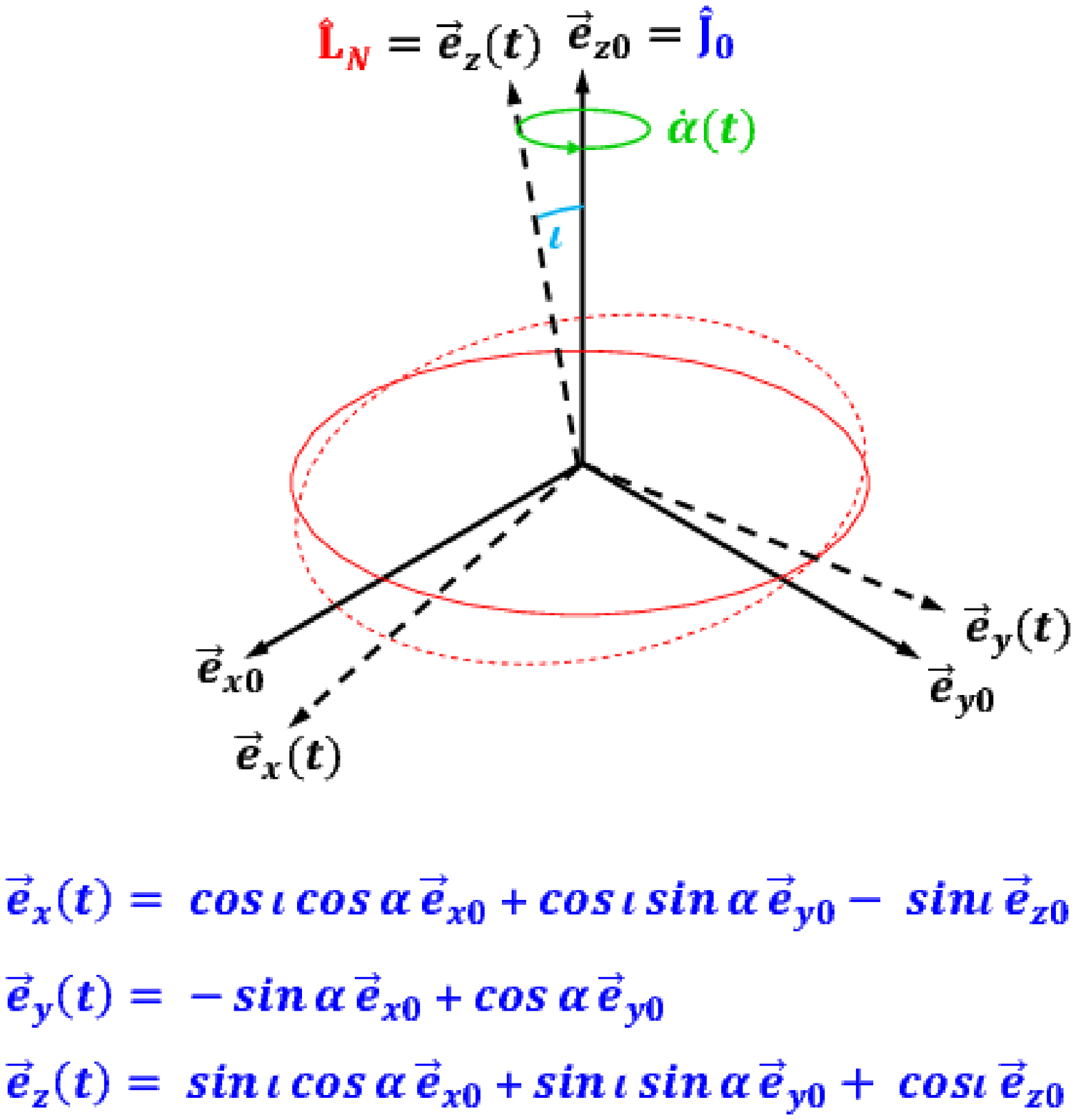}

The total angular momentum \begin{equation}
\mathbf{J}=\mathbf{J}_{0}=\mathbf{L}_{N}+\mathbf{S}\label{eq:26}\end{equation}
is assumed to be conserved (as radiation reaction is not included
throughout our analysis) and is directed along the fixed z-axis in
our analysis. We set $\hat{\mathbf{J}}_{0}=\vec{e}_{z0}$ and $\hat{\mathbf{L}}_{N}=\vec{e}_{z}(t)$,
and via the relation of Eq.~(\ref{eq:25}), Eq.~(\ref{eq:26}) may
be rewritten, \begin{equation}
J_{0}\left(-\sin\iota\vec{e}_{x}(t)+\cos\iota\hat{\mathbf{L}}_{N}\right)=L_{N}\hat{\mathbf{L}}_{N}+S\hat{\mathbf{S}},\label{eq:27}\end{equation}
where $J_{0}$, $L_{N}$ and $S$ denote the magnitudes of $\mathbf{J}_{0}$,
$\mathbf{L}_{N}$ and $\mathbf{S}$, respectively. From this one defines
\begin{equation}
\cos\beta\equiv\hat{\mathbf{S}}\cdot\hat{\mathbf{L}}_{N}=\frac{J_{0}\cos\iota-L_{N}}{S},\label{eq:28}\end{equation}
then finds easily \begin{equation}
\cos\iota=\frac{L_{N}+S\cos\beta}{J_{0}}.\label{eq:29}\end{equation}
Also, from Eqs.~(\ref{eq:26}) and (\ref{eq:28}) we have \begin{equation}
J_{0}^{2}=L_{N}^{2}+2L_{N}S\cos\beta+S^{2}.\label{eq:30}\end{equation}

\subsubsection{The angle of inclination}

According to Ref.\citet{Arun et al:2009}, in the limit $S\ll L$
the angle $\iota$ can be considered a 0.5 PN correction, and taking
advantage of this, we can reduce the lengthy expression for the GW
polarizations to a much more compact form. Within this scheme, one
finds out of Eqs.~(\ref{eq:29}) and (\ref{eq:30})\begin{eqnarray}
\cos\iota & = & \frac{1+\cos\beta\left(\frac{S}{L_{N}}\right)}{\sqrt{1+2\cos\beta\left(\frac{S}{L_{N}}\right)+\left(\frac{S}{L_{N}}\right)^{2}}}\nonumber \\
 & \begin{array}{c}
\\\overrightarrow{\,\,\,{}^{S\ll L_{N}}\,\,\,}\end{array} & 1-\frac{1}{2}\sin^{2}\beta\left(\frac{S}{L_{N}}\right)^{2}+\mathcal{O}\left(\left(S/L_{N}\right)^{3}\right).\label{eq:31}\end{eqnarray}
Now, in order to specify the angle $\iota$, we need to determine
all the quantities involved in Eq.~(\ref{eq:31}). First, the magnitude
of the Newtonian angular momentum can be replaced by the leading order
expression in $\omega_{\mathrm{orb}}$, \begin{equation}
L_{N}=\nu M^{5/3}\omega_{\mathrm{orb}}^{-1/3}=\frac{1}{4}M^{2}v^{-1},\label{eq:32}\end{equation}
where the latter expression is obtained using Eqs.~(\ref{eq:8})
and (\ref{eq:14}). Next, from Eq.~(\ref{eq:22}), one finds that
the magnitude of the total spin angular momentum $S$ is constant
(so is $\chi_{s}$) for our equal mass binary: \begin{equation}
\frac{1}{2}\frac{d}{dt}\left(S^{2}\right)=\mathbf{S}\cdot\mathbf{\dot{S}}=\mathbf{S}\cdot(\mathbf{\Omega}\times\mathbf{S})=0.\label{eq:33}\end{equation}
From this fact and from Eq.~(\ref{eq:28}) together with Eqs.~(\ref{eq:21}),
(\ref{eq:22}) and (\ref{eq:23}), one also finds that the angle $\beta$
is constant for the equal mass binary:\begin{equation}
\frac{d}{dt}\left(\cos\beta\right)=\dot{\hat{\mathbf{S}}}\cdot\hat{\mathbf{L}}_{N}+\hat{\mathbf{S}}\cdot\dot{\hat{\mathbf{L}}}_{N}=\frac{\mathbf{\dot{S}}\cdot\hat{\mathbf{L}}_{N}+\mathbf{S}\cdot\dot{\hat{\mathbf{L}}}_{N}}{S}=-\frac{4v}{M^{2}}\frac{\mathbf{S}\cdot\dot{\mathbf{S}}}{S}=0.\label{eq:34}\end{equation}
Then via Eqs.~(\ref{eq:31}), (\ref{eq:32}), (\ref{eq:33}) and
(\ref{eq:34}), one can finally specify \begin{equation}
\cos\iota=1-\frac{8S^{2}\sin^{2}\beta}{M^{4}}v^{2}+\mathcal{O}\left(v^{3}\right)\;\;\;\;\;\;\;\;\textrm{if }S\ll L.\label{eq:35}\end{equation}
From this we may also infer\begin{equation}
\iota\approx\frac{4S\sin\beta}{M^{2}}v\;\;\;\;\;\;\;\;\textrm{if }S\ll L.\label{eq:36}\end{equation}

\subsubsection{The orbital frequency}

The equation for the orbital frequency, Eq.~(\ref{eq:24}), can be
integrated in a straightforward manner. First, we rewrite it using
Eqs.~(\ref{eq:8}) and (\ref{eq:28}), \begin{equation}
\dot{v}=\frac{8}{5M}v^{9}\left[1-\frac{487}{168}v^{2}+\left(4\pi-\frac{47}{3}\frac{S\cos\beta}{M^{2}}\right)v^{3}\right].\label{eq:37}\end{equation}
Integrating this with respect to $v$, we obtain\begin{equation}
\Theta=\frac{1}{\left(2v\right)^{8}}\left[1+\frac{487}{126}v^{2}+\left(-\frac{32\pi}{5}+\frac{376}{15}\frac{S\cos\beta}{M^{2}}\right)v^{3}\right],\label{eq:38}\end{equation}
where\begin{equation}
\Theta\equiv\frac{t_{\mathrm{c}}-t}{20M},\label{eq:39}\end{equation}
and $t_{\mathrm{c}}$ denotes the instance of coalescence, at which
the frequency tends to infinity (the Post-Newtonian method breaks
down well before this point)\citet{Blanchet:2002-3}. Now, one can
invert Eq.~(\ref{eq:38}) and solve it for $v$,\begin{equation}
v=\frac{1}{2}\Theta^{-1/8}\left[1+\frac{487}{4032}\Theta^{-1/4}+\left(-\frac{\pi}{10}+\frac{47}{120}\frac{S\cos\beta}{M^{2}}\right)\Theta^{-3/8}+\mathcal{O}\left(\frac{1}{\Theta^{1/2}}\right)\right].\label{eq:40}\end{equation}
Via Eq.~(\ref{eq:8}), one finds further\begin{equation}
\omega_{\mathrm{orb}}=\frac{v^{3}}{M}=\frac{1}{8M}\Theta^{-3/8}\left[1+\frac{487}{1344}\Theta^{-1/4}+\left(-\frac{3\pi}{10}+\frac{47}{40}\frac{S\cos\beta}{M^{2}}\right)\Theta^{-3/8}+\mathcal{O}\left(\frac{1}{\Theta^{1/2}}\right)\right].\label{eq:41}\end{equation}

\subsubsection{The precession frequency }

The precession frequency $\dot{\alpha}$ also needs to be determined.
Combining Eqs.~(\ref{eq:22}) and (\ref{eq:23}), and via Eqs.~(\ref{eq:21}),
(\ref{eq:27}) and (\ref{eq:32}) we find \begin{equation}
-\frac{7}{8}J_{0}M^{2/3}\omega_{\mathrm{orb}}^{5/3}\sin\iota\vec{e}_{y}(t)=-\frac{1}{4}M^{5/3}\omega_{\mathrm{orb}}^{-1/3}\left[\dot{\iota}\vec{e}_{x}(t)+\dot{\alpha}\sin\iota\vec{e}_{y}(t)\right].\label{eq:42}\end{equation}
Here, the $\vec{e}_{x}(t)$ term being absent from the left-hand side
can be justified by comparison of $\dot{\iota}$ and $\dot{\alpha}$
on the right-hand side of Eq.~(\ref{eq:42}). To do so, we compare
the $\vec{e}_{y}(t)$ terms on the both sides first to find \begin{equation}
\dot{\alpha}=\frac{7J_{0}}{2M}\omega_{\mathrm{orb}}^{2}=\frac{7J_{0}}{2M^{3}}v^{6}.\label{eq:43}\end{equation}
Combining this with Eqs.~(\ref{eq:30}) and (\ref{eq:32}), we have
\begin{equation}
\dot{\alpha}=\frac{7}{8M}v^{5}\left[1+\frac{4S\cos\beta}{M^{2}}v+\mathcal{O}\left(v^{2}\right)\right],\label{eq:J1}\end{equation}
in the limit $S\ll L$. Now, from Eqs.~(\ref{eq:36}) and (\ref{eq:37})
we find \begin{equation}
\dot{\iota}\approx\frac{32S\sin\beta}{5M^{3}}v^{9}.\label{eq:44}\end{equation}
Evidently, this quantity is much smaller than $\dot{\alpha}$ therefore
can be ignored in our analysis. By Eq.~(\ref{eq:40}) we specify
Eq.~(\ref{eq:J1}) further\begin{equation}
\dot{\alpha}=\frac{7}{256M}\Theta^{-5/8}\left[1+\frac{2S\cos\beta}{M^{2}}\Theta^{-1/8}+\mathcal{O}\left(\frac{1}{\Theta^{1/4}}\right)\right].\label{eq:45}\end{equation}

\subsubsection{The spin vectors}

As shown by Eq.~(\ref{eq:33}) above, $S$, the magnitude of the
total spin $\mathbf{S}$ is constant. So is $\chi_{s}$ due to Eq.~(\ref{eq:17}). 

First, the components of $\mathbf{S}$ in the basis $\{\vec{e}_{x}(t),\vec{e}_{y}(t),\vec{e}_{z}(t)(=\hat{\mathbf{L}}_{N})\}$
can be found by solving Eq.~(\ref{eq:22}). To do so, we insert \begin{equation}
\mathbf{S}=S^{x(t)}\vec{e}_{x}(t)+S^{y(t)}\vec{e}_{y}(t)+S^{z(t)}\hat{\mathbf{L}}_{N}\label{eq:S1}\end{equation}
into the equation. We have then, \begin{eqnarray}
S^{x(t)} & = & \frac{-S^{z(t)}\sin\iota\dot{\alpha}}{\cos\iota\dot{\alpha}-\frac{7}{8}M^{2/3}\omega_{\mathrm{orb}}^{5/3}},\label{eq:S2}\\
S^{y(t)} & = & \frac{S^{z(t)}\dot{\iota}}{\cos\iota\dot{\alpha}-\frac{7}{8}M^{2/3}\omega_{\mathrm{orb}}^{5/3}},\label{eq:S3}\end{eqnarray}
where the contribution from $\dot{S^{x(t)}}\vec{e}_{x}(t)+\dot{S^{y(t)}}\vec{e}_{y}(t)+\dot{S^{z(t)}}\hat{\mathbf{L}}_{N}$
of $\dot{\mathbf{S}}$ on the left-hand side of Eq.~(\ref{eq:22})
has been disregarded since its magnitude is much smaller than that
of $S^{x(t)}\dot{\vec{e}_{x}}(t)+S^{y(t)}\dot{\vec{e}_{y}}(t)+S^{z(t)}\dot{\hat{\mathbf{L}}}{}_{N}$.
From Eqs.~(\ref{eq:28}), (\ref{eq:33}) and (\ref{eq:34}) above,
we find \begin{equation}
S^{z(t)}=S\cos\beta=\textrm{const.}\label{eq:S4}\end{equation}
Plugging Eqs.~(\ref{eq:35}), (\ref{eq:J1}), (\ref{eq:44}) and
(\ref{eq:S4}) into Eqs.~(\ref{eq:S2}) and (\ref{eq:S3}), and using
Eq.~(\ref{eq:8}), we obtain \begin{eqnarray}
S^{x(t)} & \approx & -S\sin\beta,\label{eq:S5}\\
S^{y(t)} & \approx & \frac{64}{35}S\sin\beta v^{3}\approx0.\label{eq:S6}\end{eqnarray}
Then via Eq.~(\ref{eq:25}) $\left[S^{x(t)},S^{y(t)},S^{z(t)}\right]$
transforms into $\left[S^{x},S^{y},S^{z}\right]$ in the basis $\{\vec{e}_{x0},\vec{e}_{y0},\vec{e}_{z0}\}$:
\begin{eqnarray}
S^{x} & \approx & S(\cos\beta\sin\iota-\sin\beta\cos\iota)\cos\alpha\approx-S\sin\beta\cos\alpha,\label{eq:S7}\\
S^{y} & \approx & S(\cos\beta\sin\iota-\sin\beta\cos\iota)\sin\alpha\approx-S\sin\beta\sin\alpha,\label{eq:S8}\\
S^{z} & \approx & S(\sin\beta\sin\iota+\cos\beta\cos\iota)\approx S\cos\beta.\label{eq:S9}\end{eqnarray}
Alternatively, we may express \begin{eqnarray}
\chi_{s}^{x} & = & -\chi_{s}\sin\beta\cos\alpha,\label{eq:S10}\\
\chi_{s}^{y} & = & -\chi_{s}\sin\beta\sin\alpha,\label{eq:S11}\\
\chi_{s}^{z} & = & \chi_{s}\cos\beta.\label{eq:S12}\end{eqnarray}

We can determine the components of $\bar{\mathbf{S}}=\mathbf{S}_{1}-\mathbf{S}_{2}$
in the basis $\{\vec{e}_{x0},\vec{e}_{y0},\vec{e}_{z0}\}$ in a similar
manner. First, one can show that the $\bar{S}$, magnitude of $\bar{\mathbf{S}}$
is also constant: so is $\chi_{a}$ due to Eq.~(\ref{eq:17-1}).
Subtracting Eq.~(\ref{eq:20}) from Eq.~(\ref{eq:19}), we have\begin{equation}
\dot{\bar{\mathbf{S}}}=\mathbf{\Omega}\times\bar{\mathbf{S}}.\label{eq:S13}\end{equation}
One finds trivially then, \begin{equation}
\frac{1}{2}\frac{d}{dt}\left(\bar{S}^{2}\right)=\bar{\mathbf{S}}\cdot\dot{\bar{\mathbf{S}}}=\bar{\mathbf{S}}\cdot(\mathbf{\Omega}\times\bar{\mathbf{S}})=0.\label{eq:S14}\end{equation}
In order to determine $\left[\bar{S}^{x(t)},\bar{S}^{y(t)},\bar{S}^{z(t)}\right]$
in the basis $\{\vec{e}_{x}(t),\vec{e}_{y}(t),\hat{\mathbf{L}}_{N}\}$,
similarly we insert \begin{equation}
\bar{\mathbf{S}}=\bar{S}^{x(t)}\vec{e}_{x}(t)+\bar{S}^{y(t)}\vec{e}_{y}(t)+\bar{S}^{z(t)}\hat{\mathbf{L}}_{N}\label{eq:S15}\end{equation}
into Eq.~(\ref{eq:S13}) and find \begin{eqnarray}
\bar{S}^{x(t)} & = & \frac{-\bar{S}^{z(t)}\sin\iota\dot{\alpha}}{\cos\iota\dot{\alpha}-\frac{7}{8}M^{2/3}\omega_{\mathrm{orb}}^{5/3}},\label{eq:S16}\\
\bar{S}^{y(t)} & = & \frac{\bar{S}^{z(t)}\dot{\iota}}{\cos\iota\dot{\alpha}-\frac{7}{8}M^{2/3}\omega_{\mathrm{orb}}^{5/3}},\label{eq:S17}\end{eqnarray}
Also, we define \begin{equation}
\cos\bar{\beta}\equiv\frac{\bar{\mathbf{S}}\cdot\hat{\mathbf{L}}_{N}}{\bar{S}},\label{eq:S18}\end{equation}
and have \begin{equation}
\bar{S}^{z(t)}=\bar{S}\cos\bar{\beta}.\label{eq:S19}\end{equation}
Now, substituting Eqs.~(\ref{eq:35}), (\ref{eq:J1}), (\ref{eq:44})
and (\ref{eq:S19}) into Eqs.~(\ref{eq:S16}) and (\ref{eq:S17}),
and using Eq.~(\ref{eq:8}), we find\begin{eqnarray}
\bar{S}^{x(t)} & \approx & -\frac{\bar{S}\cos\bar{\beta}\sin\beta}{\cos\beta},\label{eq:S20}\\
\bar{S}^{y(t)} & \approx & \frac{64}{35}\frac{S\cos\bar{\beta}\sin\beta}{\cos\beta}v^{3}\approx0.\label{eq:S21}\end{eqnarray}
However, the magnitude $\bar{S}$ calculated by means of Eqs.~(\ref{eq:S19}),
(\ref{eq:S20}) and (\ref{eq:S21}) shows \begin{equation}
\bar{S}\approx\bar{S}\frac{\cos\bar{\beta}}{\cos\beta},\label{eq:S22}\end{equation}
and we easily find \begin{equation}
\cos\bar{\beta}\approx\cos\beta=\textrm{const.}\label{eq:S23}\end{equation}
Then we may write \begin{eqnarray}
\bar{S}^{x(t)} & \approx & -\bar{S}\sin\beta,\label{eq:S24}\\
\bar{S}^{y(t)} & \approx & 0,\label{eq:S25}\\
\bar{S}^{z(t)} & \approx & \bar{S}\cos\beta.\label{eq:S26}\end{eqnarray}
The rest of procedure to transform $\left[\bar{S}^{x(t)},\bar{S}^{y(t)},\bar{S}^{z(t)}\right]$
into $\left[\bar{S}^{x},\bar{S}^{y},\bar{S}^{z}\right]$ in the basis
$\{\vec{e}_{x0},\vec{e}_{y0},\vec{e}_{z0}\}$ is the same as the above.
We finally have \begin{eqnarray}
\bar{S}^{x} & \approx & -\bar{S}\sin\beta\cos\alpha,\label{eq:S27}\\
\bar{S}^{y} & \approx & -\bar{S}\sin\beta\sin\alpha,\label{eq:S28}\\
\bar{S}^{z} & \approx & \bar{S}\cos\beta.\label{eq:S29}\end{eqnarray}
Or alternatively,\begin{eqnarray}
\chi_{a}^{x} & \approx & -\chi_{a}\sin\beta\cos\alpha,\label{eq:S30}\\
\chi_{a}^{y} & \approx & -\chi_{a}\sin\beta\sin\alpha,\label{eq:S31}\\
\chi_{a}^{z} & \approx & \chi_{a}\cos\beta.\label{eq:S32}\end{eqnarray}

\subsection{Determination of the total phase and the amplitude factor}

\subsubsection{The total phase}

\emph{Arun et. al}\citet{Arun et al:2009} defines the orbital separation
vector $\hat{\mathbf{n}}(t)$, which is set to lie along $\vec{e}_{x}(t)$
at initial time, i.e., $\hat{\mathbf{n}}(t=0)=\vec{e}_{x}(t=0)$,
and rotates on the plane spanned by $\vec{e}_{x}(t)$ and $\vec{e}_{y}(t)$
by the cumulative angle $\Phi(t)$. Then one may write down the following
two orthogonal vectors:\begin{eqnarray}
\hat{\mathbf{n}}(t) & = & \vec{e}_{x}(t)\cos\Phi(t)+\vec{e}_{y}(t)\sin\Phi(t),\label{eq:46}\\
\hat{\mathbf{\lambda}}(t) & = & -\vec{e}_{x}(t)\sin\Phi(t)+\vec{e}_{y}(t)\cos\Phi(t).\label{eq:47}\end{eqnarray}
We see that $\Phi(t)$ is the phase measuring how $\hat{\mathbf{n}}(t)$
has rotated relative to the vector $\vec{e}_{x}(t)$. For our precessing
binary, however, $\vec{e}_{x}(t)$ is itself rotating about $\hat{\mathbf{L}}_{N}$($=\vec{e}_{z}(t)$),
which is associated with the angles $\iota$ and $\alpha$ (see Eq.~(\ref{eq:25})).
Therefore the total rotation of $\hat{\mathbf{n}}(t)$ about $\hat{\mathbf{L}}_{N}$
should be a combination of a rotation of $\hat{\mathbf{n}}(t)$ with
respect to the comoving basis $\vec{e}_{x}(t)$ and $\vec{e}_{y}(t)$,
which is parametrized by $\Phi(t)$ and a rotation of the basis due
to the precession, which is parametrized by $\iota$ and $\alpha$.
Ref.\citet{Arun et al:2009} shows this in the following way. In the
basis $\{\hat{\mathbf{n}},\hat{\mathbf{\lambda}},\hat{\mathbf{L}}_{N}\}$,
we introduce an {}``orbital-like'' frequency $\omega_{\mathrm{orb}}$,
which is defined as $\omega_{\mathrm{orb}}=(\mathbf{v}\cdot\hat{\mathbf{\lambda}})/r$.
Then one may write\begin{equation}
\mathbf{v}=\dot{r}\hat{\mathbf{n}}+r\omega_{\mathrm{orb}}\hat{\mathbf{\lambda}}.\label{eq:48}\end{equation}
Now, by means of Eqs.~(\ref{eq:25}), (\ref{eq:46}) and (\ref{eq:47})
one can show \begin{equation}
\dot{\hat{\mathbf{n}}}=\left(\dot{\Phi}+\cos\iota\dot{\alpha}\right)\hat{\lambda}-\left(\dot{\iota}\cos\Phi+\sin\iota\sin\Phi\dot{\alpha}\right)\hat{\mathbf{L}}_{N}.\label{eq:49}\end{equation}
However, by imposing $\hat{\mathbf{L}}_{N}=\hat{\mathbf{n}}\times\mathbf{v}/\left|\hat{\mathbf{n}}\times\mathbf{v}\right|=\hat{\mathbf{n}}\times\dot{\hat{\mathbf{n}}}/\left|\hat{\mathbf{n}}\times\dot{\hat{\mathbf{n}}}\right|$,
one finds that the term proportional to $\hat{\mathbf{L}}_{N}$ in
Eq.~(\ref{eq:49}) must be zero. Thus, we have $\dot{\hat{\mathbf{n}}}\equiv(\mathbf{v}\cdot\hat{\mathbf{\lambda}})\hat{\mathbf{\lambda}}$,
and by identifying this with Eq.~(\ref{eq:49}) via Eq.~(\ref{eq:48})
and , we obtain\begin{equation}
\omega_{\mathrm{orb}}=\dot{\Phi}+\cos\iota\dot{\alpha},\label{eq:50}\end{equation}
which may be now interpreted as the angular velocity with which $\hat{\mathbf{n}}$
rotates about $\hat{\mathbf{L}}_{N}$. The phase $\Phi(t)$ is then
the integral\begin{equation}
\Phi(t)=\int_{0}^{t}\left[\omega_{\mathrm{orb}}(t')-\cos\iota(t')\dot{\alpha}(t')\right]dt'.\label{eq:51}\end{equation}
By plugging Eqs.~(\ref{eq:35}), (\ref{eq:41}) and (\ref{eq:45})
into Eq.~(\ref{eq:51}), and using Eqs.~(\ref{eq:39}) and (\ref{eq:40}),
one finally computes $\Phi(t)$ in the limit $S\ll L$:\begin{equation}
\Phi(t)=-4\Theta^{5/8}\left[1+\left(\frac{2435}{4032}-p_{\mathrm{S}}\frac{35}{96}\right)\Theta^{-1/4}+\left(-\frac{3\pi}{4}+p_{\mathrm{S}}\frac{59\chi_{s}\cos\beta}{64}\right)\Theta^{-3/8}+\mathcal{O}\left(\frac{1}{\Theta^{1/2}}\right)\right],\label{eq:52}\end{equation}
where $p_{\mathrm{S}}=0$ for the non-spinning case and $p_{\mathrm{S}}=1$
for the spinning case, and $S$ has been replaced by $\frac{1}{2}M^{2}\chi_{s}$
via Eq.~(\ref{eq:17}).

Also, we need a separate expression for the phase of precession. Integrating
Eq.~(\ref{eq:45}) with respect to $t$ via Eq.~(\ref{eq:39}),
and using Eq.~(\ref{eq:17}), we obtain\begin{equation}
\alpha(t)=-p_{\mathrm{S}}\frac{35}{24}\Theta^{3/8}\left[1+\frac{3\chi_{s}\cos\beta}{2}\Theta^{-1/8}+\mathcal{O}\left(\frac{1}{\Theta^{1/4}}\right)\right],\label{eq:alpha}\end{equation}
where $p_{\mathrm{S}}=0$ for the non-spinning case and $p_{\mathrm{S}}=1$
for the spinning case.

\subsubsection{The amplitude factor}

Ref.\citet{Arun et al:2009} gives the expressions for the waveform
polarizations in the following form:\begin{eqnarray}
h_{+,\times} & = & \frac{2M\nu v^{2}}{D_{L}}\left[H_{+,\times}^{(0)}+H_{+,\times}^{(1/2)}+H_{+,\times}^{(1/2,\mathrm{SO})}+H_{+,\times}^{(1)}+H_{+,\times}^{(1,\mathrm{SO})}\right.\nonumber \\
 &  & \left.+H_{+,\times}^{(3/2)}+H_{+,\times}^{(3/2,\mathrm{SO})}\right].\label{eq:53}\end{eqnarray}
In our equal-mass case ($\nu=1/4$), the amplitude factor for each
PN group will then be determined by means of Eq.~(\ref{eq:40}) \begin{eqnarray}
F^{(0)} & = & \frac{Mv^{2}}{2D_{L}}=\frac{M}{8D_{L}}\Theta^{-1/4},\label{eq:54}\\
F^{(1/2)} & = & \frac{Mv^{3}}{2D_{L}}=\frac{M}{16D_{L}}\Theta^{-3/8},\label{eq:55}\\
F^{(1)} & = & \frac{Mv^{4}}{2D_{L}}=\frac{M}{32D_{L}}\Theta^{-1/2}\left(1+\frac{487}{1008}\Theta^{-1/4}\right),\label{eq:56}\\
F^{(3/2)} & = & \frac{Mv^{5}}{2D_{L}}=\frac{M}{64D_{L}}\Theta^{-5/8}\left[1+\frac{2435}{4032}\Theta^{-1/4}+\left(-\frac{\pi}{2}+p_{\mathrm{S}}\frac{47\chi_{s}\cos\beta}{48}\right)\Theta^{-3/8}\right],\label{eq:57}\end{eqnarray}
where $p_{\mathrm{S}}=0$ for the non-spinning case and $p_{\mathrm{S}}=1$
for the spinning case, and in the last equation $S$ has been replaced
by $\frac{1}{2}M^{2}\chi_{s}$ via Eq.~(\ref{eq:17}).

\section*{SUMMARY}

\textbf{$ $}\\
\textbf{1. The total phase}

\[
\Phi(t)=-4\Theta^{5/8}\left[1+\left(\frac{2435}{4032}-p_{\mathrm{S}}\frac{35}{96}\right)\Theta^{-1/4}+\left(-\frac{3\pi}{4}+p_{\mathrm{S}}\frac{59\chi_{s}\cos\beta}{64}\right)\Theta^{-3/8}+\mathcal{O}\left(\frac{1}{\Theta^{1/2}}\right)\right]\]
\\
\textbf{2. The precession phase}

\[
\alpha(t)=-p_{\mathrm{S}}\frac{35}{24}\Theta^{3/8}\left[1+\frac{3\chi_{s}\cos\beta}{2}\Theta^{-1/8}+\mathcal{O}\left(\frac{1}{\Theta^{1/4}}\right)\right]\]
\\
\textbf{3. Inclination angle}

\[
\iota\approx2\chi_{s}\sin\beta v\;\;\;\;\;\;\;\;\textrm{if }S\ll L\]
\\
\textbf{4. The spin vectors}\[
\left[\begin{array}{l}
\chi_{s}^{x}\\
\chi_{s}^{y}\\
\chi_{s}^{z}\end{array}\right]=\left[\begin{array}{l}
-\chi_{s}\sin\beta\cos\alpha\\
-\chi_{s}\sin\beta\sin\alpha\\
\chi_{s}\cos\beta\end{array}\right]\]
\[
\left[\begin{array}{l}
\chi_{a}^{x}\\
\chi_{a}^{y}\\
\chi_{a}^{z}\end{array}\right]=\left[\begin{array}{l}
-\chi_{a}\sin\beta\cos\alpha\\
-\chi_{a}\sin\beta\sin\alpha\\
\chi_{a}\cos\beta\end{array}\right]\]
\textbf{}\\
\textbf{5. The amplitude factors}

\begin{eqnarray*}
F^{(0)} & = & \frac{Mv^{2}}{2D_{L}}=\frac{M}{8D_{L}}\Theta^{-1/4}\\
F^{(1/2)} & = & \frac{Mv^{3}}{2D_{L}}=\frac{M}{16D_{L}}\Theta^{-3/8}\\
F^{(1)} & = & \frac{Mv^{4}}{2D_{L}}=\frac{M}{32D_{L}}\Theta^{-1/2}\left(1+\frac{487}{1008}\Theta^{-1/4}\right)\\
F^{(3/2)} & = & \frac{Mv^{5}}{2D_{L}}=\frac{M}{64D_{L}}\Theta^{-5/8}\left[1+\frac{2435}{4032}\Theta^{-1/4}+\left(-\frac{\pi}{2}+p_{\mathrm{S}}\frac{47\chi_{s}\cos\beta}{48}\right)\Theta^{-3/8}\right]\end{eqnarray*}
\\
Above, $p_{\mathrm{S}}=0$ for the non-spinning case and $p_{\mathrm{S}}=1$
for the spinning case.\\
$\chi_{s}\equiv2S/M^{2}=\textrm{constant}$ and $\chi_{a}\equiv2\bar{S}/M^{2}=\textrm{constant}$;
$\mathbf{S}\equiv\mathbf{S}_{1}+\mathbf{S}_{2}$, $\bar{\mathbf{S}}\equiv\mathbf{S}_{1}-\mathbf{S}_{2}$
$M\equiv M_{1}+M_{2}$. \\
$\beta\equiv\cos^{-1}\left(\frac{\mathbf{L}_{N}\cdot\mathbf{S}}{L_{N}S}\right)=\cos^{-1}\left(\frac{\mathbf{L}_{N}\cdot\bar{\mathbf{S}}}{L_{N}\bar{S}}\right)=\textrm{constant}$.\\

\end{document}